\documentclass[conference]{IEEEtran}
\IEEEoverridecommandlockouts
\usepackage{balance}
\usepackage{cite}
\usepackage{amsmath,amssymb,amsfonts}
\usepackage{algorithmic}
\usepackage{graphicx}
\usepackage{textcomp}
\usepackage{xcolor}
\def\BibTeX{{\rm B\kern-.05em{\sc i\kern-.025em b}\kern-.08em
    T\kern-.1667em\lower.7ex\hbox{E}\kern-.125emX}}
    
\usepackage{mathtools} 
\DeclareMathOperator*{\argmin}{argmin}

\usepackage{fancyhdr}
\usepackage{kantlipsum}
\fancypagestyle{plain}{
\fancyhf{}
\fancyhead[C]{Conference on \LaTeX} 

}
\usepackage{eso-pic}
\usepackage{hyperref}

\begin{document}

\AddToShipoutPictureBG*{
\AtPageUpperLeft{
\setlength\unitlength{1in}
\hspace*{\dimexpr0.5\paperwidth\relax}
\makebox(0,-0.75)[c]{\textbf{2022 IEEE/ACM International Conference on Advances in Social Networks Analysis and Mining (ASONAM)}}}}

\title{Causal Analysis on the Anchor Store Effect in a Location-based Social Network}

\author{
\IEEEauthorblockN{
Anish K. Vallapuram\IEEEauthorrefmark{1}, Young D. Kwon\IEEEauthorrefmark{2}, Lik-Hang Lee\IEEEauthorrefmark{3}, Fengli Xu\IEEEauthorrefmark{4},  and Pan Hui\IEEEauthorrefmark{1}
}
\IEEEauthorblockA{\IEEEauthorrefmark{1}Hong Kong University of Science and Technology, Hong Kong SAR}
\IEEEauthorblockA{\IEEEauthorrefmark{2}University of Cambridge, United Kingdom}
\IEEEauthorblockA{\IEEEauthorrefmark{3}Korea Advanced Institute of Science and Technology, South Korea}
\IEEEauthorblockA{\IEEEauthorrefmark{4}University of Chicago, United States}
Email: akvallapuram@connect.ust.hk
}


\IEEEoverridecommandlockouts
\IEEEpubid{
    \parbox{\textwidth}{
        \makebox[\textwidth][t]{
            IEEE/ACM ASONAM 2022, November 10-13, 2022}
        \makebox[\textwidth][t]{
            \url{http://dx.doi.org/10.1145/XXXXXXX.XXXXXXX}} 
        \makebox[\textwidth][t]{
        978-1-6654-5661-6/22/\$31.00~\copyright\space2022 IEEE} 
    }
}

\maketitle

\IEEEpubidadjcol

\begin{abstract}
    A particular phenomenon of interest in Retail Economics is the spillover effect of anchor stores (specific stores with a reputable brand) to non-anchor stores in terms of customer traffic. Prior works in this area rely on small and survey-based datasets that are often confidential or expensive to collect on a large scale. Also, very few works study the underlying causal mechanisms between factors that underpin the spillover effect.
     In this work, we analyze the causal relationship between anchor stores and customer traffic to non-anchor stores and employ a propensity score matching framework to investigate this effect more efficiently.
     First of all, to demonstrate the effect, we leverage open and mobile data from London Datastore and Location-Based Social Networks (LBSNs) such as Foursquare. We then perform a large-scale empirical analysis of customer visit patterns from anchor stores to non-anchor stores (e.g., non-chain restaurants) located in the Greater London area as a case study. By studying over 600 neighbourhoods in the Greater London area, we find that anchor stores cause a 14.2-26.5\% increase in customer traffic for the non-anchor stores reinforcing the established economic theory 
     Moreover, we evaluate the efficiency of our methodology by studying the confounder balance, dose difference and performance of the matching framework on synthetic data. Through this work, we point decision-makers in the retail industry to a more systematic approach to estimate the anchor store effect and pave the way for further research to discover more complex causal relationships underlying this effect with open data.
\end{abstract}

\section{Introduction}
\label{sec:Introduction}

Retail store co-location is a widely studied problem in the field of Economics~\cite{anchor_stores, vitorino, Eppli1994TheEO, Damian}. A particular phenomenon of interest is the \emph{anchor store effect}. The idea of an anchor store originates from large department stores which cover a wide range of products which causes a spillover of shopper footfall within its vicinity (i.e., to non-anchor stores). This can occur due to multitudinous factors including but not limited to the brand name, consumer awareness of the product prices and their reduced commute costs per product category. Anchor stores are advantageous to owners of small businesses and malls alike as they can leverage the footfall from the anchor stores' brand name. Motivated by this, mall owners offer discounted rents to anchor stores for being their tenants~\cite{Gould}.

Previous studies mainly focus on the effect of large anchor stores in malls on the sales and customer traffic of small non-anchor tenants~\cite{Eppli1994TheEO}. Several works~\cite{Benjamin1990, Sirmans} have investigated correlations between the rents paid by both anchor and non-anchor tenants to the sales. Some other works~\cite{FINN, anderson1985Assoc.} have conducted surveys on the prospect that customers in a neighbourhood will visit a mall given the brand reputation of its anchor tenants. However, these works only consider mall-related variables which are insufficient for the study as later empirical studies \cite{davidson2000retail, artz2006analyzing, hernandez2003impact} demonstrate the effect of the entrance of anchor stores on the retail trade at the town level. Furthermore, Daunfeldt et al. \cite{ikea} expostulate that these works employ descriptive statistics to quantify the effects and do not draw causal conclusions. Still, these works examine the impact of anchor stores on retail sales, profit margins and employment. The data for these variables is either under non-disclosure agreements or cost-ineffective to acquire at a large scale. This limits the studies to a specific retail brand, ignoring the other anchor stores in the vicinity.

\begin{figure*}[t!]
    \centering
    \includegraphics[scale=0.38]{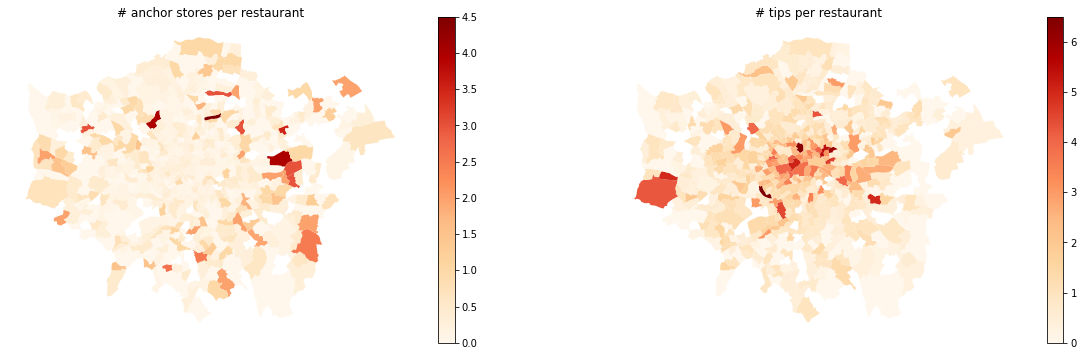}
    \caption{Our problem formulation --  studying the anchor store effects on non-chain restaurants. The number of anchor stores per restaurant (left) and the number of tips per restaurant (right) are shown for each neighbourhood in the Greater London area. Data in 2012 are depicted, with the darker colour representing a higher count of stores/tips per restaurant or vice versa.}
    \label{fig:treat_out_map}
\end{figure*}

In this paper, we instead leverage publicly available user mobility data generated by the location-based social networks (LBSNs), \textit{Foursquare}, and census data \textit{London Datastore} which covers observational data for 625 neighbourhoods across Greater London over three years.  We then estimate the causal impact (i.e., change in customer traffic) of the presence of anchor stores (e.g. Tesco, Sainsbury's, Waitrose, Lidl) on non-anchor stores located in the Greater London area as a case study (Figure~\ref{fig:treat_out_map}). In our initial causal analysis, we focus on non-chain restaurants as a representative example of non-anchor stores to narrow down the scope of this study. Yet, our method is generalisable to any other type of non-anchor store. Furthermore, we go beyond the mall-related variables and consider several neighbourhood characteristics that significantly impact customer traffic to retail stores~\cite{Sirmans}. 
Additionally, prior works either ignore these variables or fail to consider their interdependence leading to biased conclusions on the effect. 
In contrast, our work mitigates the aforementioned issue by employing a causality analysis framework that handles the interdependence between variables before performing the inference and drawing conclusions. 
To the best of our knowledge, this is the first work to leverage data from LBSNs and employ causal inference to demonstrate the effect of anchor stores on increased customer traffic to neighbouring non-anchor stores.

Overall, the major contributions of this paper are:
\begin{itemize}
    \item To the best of our knowledge, this is the first work to employ observational data from an LBSN to perform a large-scale causal study on this effect with over 1,583 anchor stores and 7,694 non-anchor restaurants in the Greater London area by observing customer traffic over three years.
    \item Based on a propensity score matching framework, we study the effectiveness of the anchor store effect by considering neighbourhood characteristics such as population density, disposable income, and transport accessibility to further mitigate their confounding effect on customer traffic to anchor stores and retail restaurants.  
    \item Our results consistently demonstrate a 14.2-26.5\% increase in customer traffic in non-anchor stores with more neighbouring anchor stores, reinforcing the well-known economic concept. We further evaluate the significance of these results through several metrics and performance on synthetic data with known causal effects.
    \item Based on the findings of this work on the causal relation between anchor stores and retail restaurants, we elaborate on how our findings are useful to various beneficiaries.
\end{itemize}


\section{Related Work}
\label{sec:Related_Work}

We will first review previous works on anchor store effects and then delve into the background of causality analysis.

\textbf{Anchor Store Effect.} 
Several works have investigated the effect of anchor stores in malls on the sales and customer traffic to non-anchor tenants~\cite{Eppli1994TheEO}. 
\cite{Benjamin1990, Sirmans} study this impact by finding the correlation between the rents paid by anchor stores and non-anchor stores and \cite{Damian} perform regression analysis on the mall visitor traffic with respect to several independent variables such as the total area of the commercial centre, age, number of stores, number of anchor stores and their total area. 
\cite{Sirmans} suggests that variables beyond the walls of the mall also affect customer traffic. Consequently, several empirical studies \cite{davidson2000retail, artz2006analyzing, hernandez2003impact} demonstrated the effect of the entrance of anchor stores on retail trade at the town level. 
For example, Davidson et al. \cite{davidson2000retail} concluded that total revenues increase in towns by 41\% in Wal-Mart entry towns versus 28\% in towns which Wal-Mart did not enter.  Yet, these works do not draw causal conclusions as they quantify the effect using descriptive statistics. Hence, Daunfeldt et al. \cite{ikea} employ a propensity score matching framework \cite{psm} to study the effect of the entrance of IKEA stores in three municipalities in Sweden and concluded a 7\% increase in revenues to incumbent retailers. 

However, these works study the impact of anchor stores on variables such as revenues, profit margins, wages, etc. which poses two key limitations: 1) the data is not available for analytics due to non-disclosure agreements, and 2) The data collection is laborious and cost-ineffective at a large scale. These works limit themselves to a specific retail brand, ignoring the other anchor stores in the vicinity. Thus, in this work, we study the impact of anchor stores on customer traffic by employing observational data from an LBSN which allows us to study the impact of 1,583 anchor stores from several retail brands on 7,694 restaurants in the Greater London area.

\textbf{Causality Analysis.} 
To ensure our findings' robustness and generalisability, we study the anchor store effect via causality analysis, a long-standing research area that lies at the heart of scientific discovery~\cite{rubin_framework}. Randomised controlled trial (RCT) is a widely accepted approach for performing this analysis, especially in medicine and biology~\cite{chalmers1981method}. However, it requires randomly assigning experiment subjects to treated and controlled groups in customised experiments, which is often expensive and infeasible. Several alternative frameworks have addressed this by estimating the causal relationships from observational data, which fall into three categories: 1) Given the underlying causal diagram of observational data, causal relation can be estimated with statistical methods, e.g., instrumental variables analysis~\cite{pearl2009causal, thistlethwaite1960regression}. However, 
such a causal diagram requires extensive expert knowledge and is difficult to validate; 
2) Under the assumption that we observe all the confounding variables, simple regression analysis can examine the causality in observational data~\cite{acemoglu2001colonial}, but these approaches are sensitive to the selected functional forms of the confounders. 
Even if we could observe all confounders, we still face the common support problem where the covariate space must align across all the subjects.
3) Another angle is to estimate the result of RCT by matching each treated subject with a controlled subject that has similar confounding variables~\cite{rosenbaum1983central}.  
In this case, the complexity of the common support problem is limited to the subjects in the matched pair.  

In this paper, 
thus design a non-randomised study that estimates the causal effect of an anchor store with a graphical propensity score matching algorithm, which falls in the third category mentioned above.

\section{Methodology}
\label{sec:Methodology}

Formally, the goal of this work is to study the effect of the presence of anchor stores within the neighbourhood of non-anchor stores on their customer visit patterns. In a causal study, the \textit{units} are the basic experimental objects belonging to a larger collection of the population which in our case are neighbourhoods in a large city. A highly relevant instance is the Greater London area where the neighbourhoods are analogous to its 625 wards\footnote{neighbourhoods, units and wards will now be used interchangeably}. The presence of the anchor stores is the \textit{treatment} denoted by $Z$, and customer visit patterns to non-anchor stores, such as non-chain restaurants, is the \textit{outcome} denoted by $Y$.

We perform causality analysis by employing Rubin's framework of potential outcomes~\cite{rubin_framework} which operates as follows. Given a set of units $U$ and a set of binary treatment levels $Z \in \{0, 1\}$, we are interested in comparing the outcome of applying each treatment level $Y_{1}(u)$ and $Y_{0}(u)$ for each unit $u$. The average treatment effect (ATE) is then $E(Y_{1}(u) - Y_{0}(u))$. Yet, the rudimentary problem of causal analysis is that both the outcomes for a unit cannot be observed, especially in our context. Therefore, a widely sought-after alternative~\cite{lu_et_al, syn_data_ex} is a matching-based approach that determines the effect of the treatment $Z$ on the outcome $Y$ by comparing pairs of units with highly similar confounder variables or simply confounders $X$. These variables describe certain fundamental characteristics of the units that have a causal effect on the treatment and outcome variables themselves. For example, customer visit patterns to non-chain restaurants (outcome) can be affected by not only the presence of the anchor stores (treatment) but also by the population density of the neighbourhood (confounder). Ignoring the effect of confounders is a fundamental fallacy in causality analysis. Once the confounders are known, a similarity of distance metric for two units can be defined based on their confounders. The final set of matching $M$ contains highly similar pairs of units $(u_{t}, u_{c})$ where the treatment values of the pair are $Z(u_{t}) = 1$ and $Z(u_{c}) = 0$ and no unit appears in more than one pair. The average treatment effect (ATE) can be computed as 

\begin{equation}
    ATE = \frac{1}{|M|}\sum_{(u_{t},u_{c}) \in M} Y(u_{t}) - Y(u_{c})
\end{equation}

The remainder of this section addresses how we adapt the matching framework to our problem based on one remark. The causal framework has been described with binary treatment levels, i.e. $Z \in (0, 1)$.  However, the treatment variable takes a range of values in observational data for our problem and thus binned into treatment levels or \textit{doses} as shown in Table~\ref{table:treatment_levels}. 

\subsection{Matching}\label{subsec:Matching}
This section details a graphical approach to match units based on confounder similarity. We first construct a fully connected graph $G = (V, E)$ where the vertices are units. Each edge $(u_{i}, u_{j}, w_{ij}) \in E$ has a weight ($w_{ij}$) determined by the distance metric based on the confounders of the units ($u_{i}, u_{j}$). An optimal matching $M \subset E$ is achieved by optimising:

\begin{equation}
    \argmin_{M} \frac{1}{|M|} \sum_{(u_{i}, u_{j}, w_{ij}) \in M} w_{ij}
    \label{equation: optimisation}
\end{equation} 

where the maximum matching must be determined by the lowest sum of edge weights. A matching $M$ is called a maximum matching if it includes the most edges from the graph, i.e. for every matching $M'$ of $G$, $|M| \geq |M'|$. In a causal experiment, outcomes are only compared between units from different treatment groups \cite{rosenbaum1983central}. This can be achieved by inducing a new graph $G' = (V, E')$ from $G$ by simply dropping all edges in $E$ where the connected units have the same treatment level, i.e.  $E' = \{(u_{i}, u_{j}, w_{ij}) \in E: z_{i} > z_{j}$.


\subsection{Propensity Score}\label{subsec:Propensity Score}

An appropriate choice of distance metric is paramount for the quality of the matching. While euclidean distances appear as obvious choices, they do not scale well with the dimensionality of the confounders and there will be non-uniform confounder distributions within the disjoint vertex sets which not ideal. An alternative metric called propensity score resolves this~\cite{psm},
which is defined as the probability that a unit receives a particular treatment level given the confounders. The propensity score function outputs the probability distribution for each treatment level, hence the dimensionality is more controlled compared to the dimensionality of the confounders. Joffe and Rosenbaum~\cite{joffe} show that the propensity score must satisfy the balancing property $P(X=X(u_{i})|e(X(u_{i})), Z=z) = P(X=X(u_{i})|e(X(u_{i})), Z=z')$ for all combinations of treatment levels $(z, z')$ where $e$ is the propensity score function. This is certainly the case when the propensity score is described using the McCullagh's ordered logit model~\cite{mccullagh} 
\begin{equation}
    log\left(\frac{P(z_i >= d)}{P(z_i < d)}\right) = \theta_{d} + \beta^{T}x_{i}
\end{equation}

where $\beta^{T}x_{i}$ is the propensity score for unit $u_{i}$and $d$ is the treatment level.  The parameter $\beta^{T}$ is estimated using maximum likelihood to give the estimate $\hat{\beta}^{T}$. The propensity score can now be incorporated into the edge weights of the graph. To further incentivise the matching algorithm to pair units with the maximised difference in treatments in the multiple treatment level case, their treatments are also integrated into the weights. The distance metric is thus 

\begin{equation}
    w_{ij} = \frac{(\hat{\beta}^{T}x_{i} - \hat{\beta}^{T}x_{j})^{2} + \epsilon}{(z_{i} - z_{j})^{2}}
\end{equation}

where $\epsilon$ is a very small positive real number put into place to mitigate the edge case where the confounders of two units are the same which invalidates the edge with a zero weight. With the graph and its edge weights now defined, the optimal matching $M$ is found using a weight version of the Edmond's Blossom algorithm~\cite{matching_alg} which optimises Equation \ref{equation: optimisation}. Finally, the average treatment effect is calculated as follows

\begin{equation}
    ATE = \frac{1}{M} \sum_{(u_{i},u_{j}, w_{ij}) \in M} \frac{y_{i} - y_{j}}{z_{i} - z_{j}}
\end{equation}

\section{Datasets}
\label{sec:Datasets}

This section contextualises our problem by describing the observation data collected and defining all the variables. 
The datasets were obtained from the following two sources:

\textbf{London Datastore\footnote{https://data.london.gov.uk/dataset/lsoa-atlas}.} The Greater London Authority maintains an open data-sharing portal that contains census data for each neighbourhood including statistics such as demographics, employment, housing and transportation. The portal also provides neighbourhood boundaries in the GeoJSON format. 

\begin{figure}[t!]
    \centering
    \includegraphics[scale=0.3]{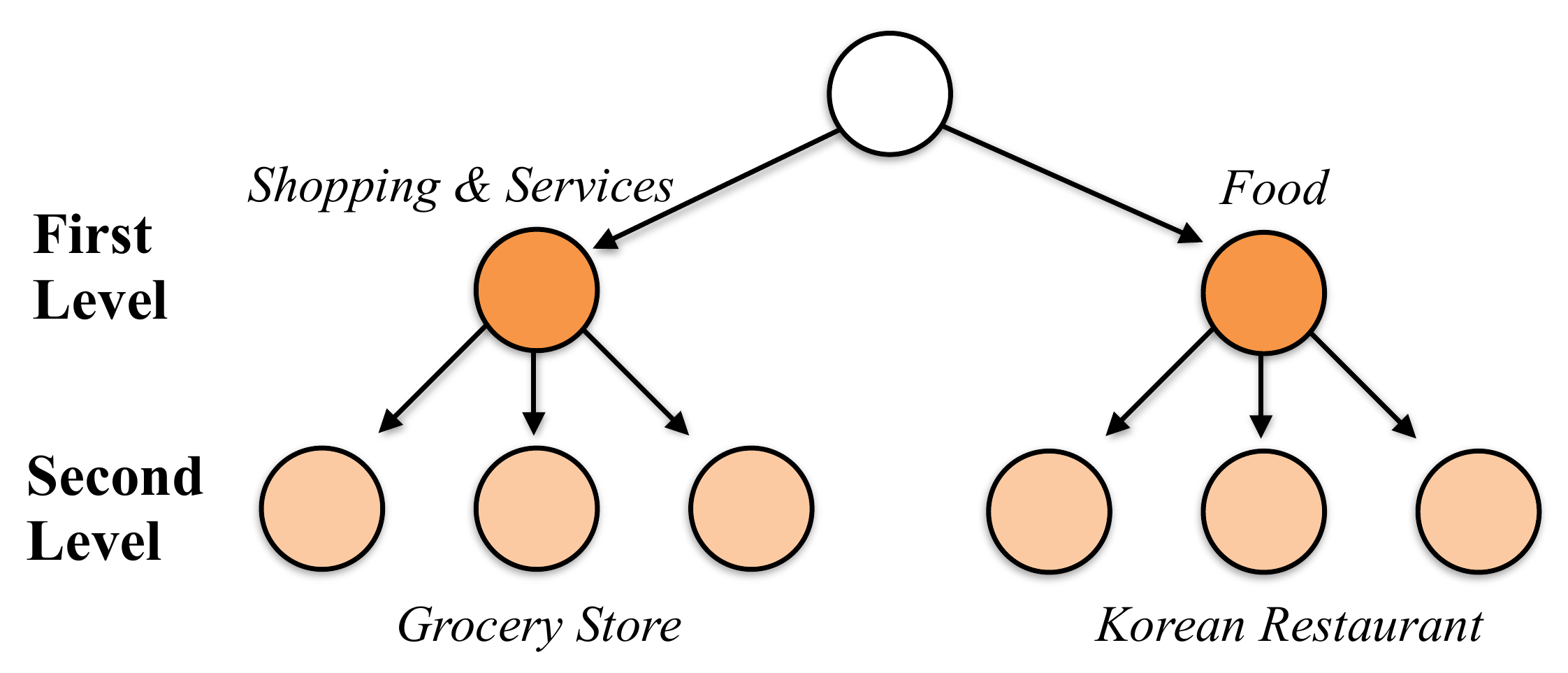}
    \caption{Hierarchy of Foursquare venue categories. Original hierarchy has five levels and more categories in each level but only two are shown.}
    \label{fig:4sq_cats}
\end{figure}

\textbf{Foursquare.} The venue and tip data for the retail stores were obtained through the LBSN platform Foursquare. Collected by Chen et al.~\cite{foursquare}, the dataset consists of 19 million tips for 13 million venues from around the world. Each venue includes information such as the venue ID, name, location and time of account creation. Venues are also assigned categories such as ``Restaurants", ``Bank", and ``Pharmacy'' and these categories have a hierarchical structure as depicted in Figure \ref{fig:4sq_cats}. Tips are timestamped texts written by users for venues. 

Since the geography of our problem is limited to the Greater London area, we initially filtered venues using the country code field to shortlist all the venues that belong to the United Kingdom. This significantly reduces the number of venues from over 13 million to 272,079. These venues were assigned to neighbourhoods based on whether any of the neighbourhood boundaries enclosed their location (given by latitude and longitude). The unassigned venues are consequently not in London and were discarded.

For reliability of results, the experiment is repeated thrice each year from 2011 to 2013. The treatment must temporally precede the outcome to measure the causal effect. Hence, the treatment and outcome are measured at the beginning and the end of the year respectively for each neighbourhood.

\subsection{Treatment}
The treatment variable must measure the presence of anchor stores which are defined as large retail chains with a brand name~\cite{anchor_stores}.
Hence, a list of global brand names was developed from all the values to the `brand' key from OpenStreetMaps~\footnote{https://taginfo.openstreetmap.org/keys/brand\#values} and the names from the `chains' field in 
Foursquare~\footnote{https://developer.foursquare.com/docs/build-with-foursquare/chains/}. The venues in Greater London are labelled as anchor stores if their name is present in the brand name list and have the first level category of `Shop \& Service'. Only anchor stores that started before the year of the experiment are counted. The treatment variable for each neighbourhood is the number of anchor stores per restaurant (hence $Z \in \mathbb{R}^{+}$). To utilise the propensity score described earlier, the treatment values are discretised into three treatment levels. Since there is no significant literature on choosing the number of treatment levels, we intuitively set it to three for two reasons (1) The number of levels dictates the dimensionality of the propensity score which must be set low; (2) Around 33\% of the neighbourhoods do not have anchor stores (i.e. treatment level is 0). Hence, we split the remainder of the neighbourhoods into two groups yielding three different treatment levels so that each level consists of roughly 33\% of the neighbourhoods. Table \ref{table:treatment_levels} shows the treatment value ranges for each treatment level. 

\subsection{Outcome}
The outcome variable must measure the customer visit patterns in restaurants in the Greater London area. Foursquare provides two types of customer activities at the venues: check-ins and tips. 
Since a prior study ~\cite{geosocial} indicated that 75\% of the check-ins are fake, we use tips following~\cite{kwon_geolifecycle_imwut19}. 
Tips provide a better proxy of customer visits because they must write a text on the platform regarding the venue which serves as a verification of their visit. The restaurants within each neighbourhood are determined using the second-level categories containing the term `Restaurant' for venues. Among restaurants that started before the year of the experiment, the number of tips posted on them during the year is counted. It is also ensured that the non-anchor restaurants do not belong to a retail chain or brand name. Finally, the outcome variable for each neighbourhood is the number of tips per restaurant. 

\begin{table}[h!]
\caption{Treatment level ranges in no. of anchor stores per restaurant in a neighbourhood}
\centering
\begin{tabular}{|l|r|r|r|}
\hline
 year / treatment level &  0 &   1 &  2 \\
\hline
 2011 & $\leq$ 0.000 & $\leq$0.250 & $>$0.250  \\
 2012 & $\leq$0.000 & $\leq$0.228 & $>$0.228 \\
 2013 & $\leq$0.000 & $\leq$0.223 & $>$0.223 \\
\hline
\end{tabular}

\label{table:treatment_levels}
\vspace{-0.3cm}
\end{table}

Table \ref{tab:top_10} summarises the top 10 second-level categories and anchor brands. Grocery stores are the most common type of anchor store which is consistent with the literature~\cite{anchor_stores, FINN} which confirmed that large departmental stores traditionally served as anchor stores. We observe that apparel stores, electronics and pharmacies are also gaining attention as anchors.

\begin{table}[h!]
    \caption{Top 10 anchor store categories and brands by percentage of stores in Greater London till 2013}
    \centering
    \begin{tabular}{|ll|ll|}
    \hline
    Category &  \% stores & Brand & \% stores \\
    \hline
    Grocery Store                 & 23.67 & Tesco & 14.01 \\
    Clothing Store                & 14.98 & Boots & 5.59 \\
    Pharmacy          & 6.43 & Sainsbury's    & 5.54 \\
    Convenience Store             & 5.33 & Co-op Food & 3.75 \\
    Bookstore                     & 4.27 & WHSmith               & 3.29 \\
    Electronics Store             & 3.08 & Iceland               & 2.54 \\
    Hardware Store                & 2.90 & Argos                 & 2.36 \\
    Shoe Store                    & 2.80 & Superdrug             & 2.25 \\
   Office Supplies  & 2.57 & Marks \& Spencer       & 2.08 \\
    Women's Store                 & 2.48 & TK Maxx               & 1.90 \\
    \hline
    \end{tabular}
    \label{tab:top_10}
\end{table}

\subsection{Confounding Variables}\label{subsec:Confounding Variables}

The selection of confounders is crucial to propensity score matching as described earlier and requires domain knowledge of the retail industry. In our context, confounding variables must be key factors for site selection that impact the success of the retail stores and restaurants. To this end, we consult several papers~\cite{key_factors, ikea, location3} and employ 15 confounders from the following categories for each neighbourhood:

\textbf{Population and Age.} Retail stores not only seek populated neighbourhoods but also target customers of a particular demographic. This is captured by three variables measuring population for age groups of (i) 0-15, (ii) 16-64 (iii) 65+. 

\textbf{Education.} Education plays an important role in determining the living standard and availability of skilled labour pool which is measured using the Average GCSE point scores.

\textbf{Housing.} The standard of living is also reflected through affordable housing accounted for by the following variables (i) median house price, (ii) number of dwellings (iii) the percentage of dwellings sold during the year. 

\textbf{Employment.} The reason for including employment-related variables is two-fold. Employment indicators not only suggest the availability of the labour pool but also potential customers with disposal income. This is reflected by the following three variables: (i) number of full-time (ii) number of part-time employees (iii) their mean income. 

\textbf{Business Accessibility.} This category measures the ability of customers to pay a visit to the retail store or restaurant. Tayeen et al.~\cite{location3} suggest that the availability of transportation access and low crime rate are relevant variables. These are measured by the average public transport accessibility score (PTAL) and the total crime rate in the datastore. 

\textbf{Tourism.} Several customers are attracted to a restaurant's neighbourhood not only because of the presence of anchor stores but also its tourism significance. According to Tayeen et al., ~\cite{location3}, the tourism significance of a neighbourhood is captured based on the counts of four types of attractions: (i) tourist locations (ii) rivers, lakes and reservoirs (iii) parks. These are measured by counting venues in Foursquare with corresponding categories in each neighbourhood of London. 




\section{Results}
\label{sec:Results}

This section presents the results of the experiments. We first match the neighbourhoods based on their propensity scores. We then evaluate the propensity score matching 
through three metrics namely:
dose difference, confounder balance and performance on a synthetic dataset~\cite{hasthanasombat_understanding_2019}. 
We finally compare the outcome of all matched pairs of neighbours and quantify the effect of anchor stores on customer traffic in Greater London. 

\subsection{Correlation Analysis}\label{subsec:Correlation Analysis}

\begin{figure}[t!]
    \centering
    \includegraphics[scale=0.4]{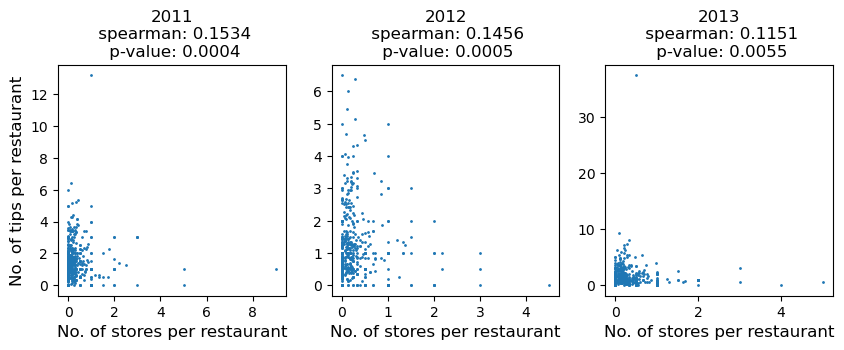}
    \caption{Correlation analysis between the treatment and outcome variables.}
    \label{fig:corr_analysis}
\end{figure}

To obtain an initial insight, we plotted the number of stores per restaurant (i.e. treatment) against the number of tips per restaurant (i.e. outcome) which is shown in figure~\ref{fig:corr_analysis}. The spearman correlation~\cite{spearman} is utilised to measure the correlation between the two variables. A small but significant positive correlation is observed between the treatment and outcome variables across all three years. Although premature, the results suggest that a higher number of anchor stores in a neighbourhood correlate with an increase in the number of tips to restaurants. Yet, the results from the following sections will establish a causal relationship between the two variables. 

\begin{table}[]
    \caption{Counts of dose levels between matched pairs of (low dose, high dose) units.}
    \label{tab:dose_diff}
    \centering
    \begin{tabular}{|c|c|c|c|}
    \hline
    year / dose pair & (0, 1) & (1, 2) & (0, 2) \\
    \hline
    2011 & 103 & 38 & 119 \\
    2012 & 93 & 66  & 121 \\
    2013 & 97 & 80 & 113 \\
    \hline
    \end{tabular}
\end{table}

\subsection{Dose Difference}\label{subsec:Dose Difference}
The dose difference exposes the efficacy of a specific aspect in the matching method: maximising the difference in treatment levels between matched pairs. This mitigates the influence of noise when calculating the average treatment effect. The matrix plots, as shown in 
Table~\ref{tab:dose_diff}, depict the distributions of doses in matched pairs for the three years (i.e., 2011, 2012 and 2013). The rows represent the treatment received by a unit at a  higher level compared to its match. The column represents the treatment level of the unit at a lower level than its counterpart. The treatment difference is two levels for 45.8\%,  43.2\% and 39.0\% of the pairs in 2011, 2012 and 2013 respectively. While the remainder of the percentages reflects the pairs with a treatment level difference of one. 

\begin{table}
\caption{The average confounder values between high and low dose units among matched pairs in 2013.}
\centering
\begin{tabular}{|p{0.35\linewidth}|r|r|r|r|}
\hline
{} &      high &       low &  t-test &  p-value \\
\hline
\% of dwellings sold            &      3.29 &      3.22 &    0.79 &     0.43 \\
Aged 0-15                                  &   2739.14 &   2647.41 &    1.32 &     0.19 \\
Aged 16-64                                 &   9526.72 &   9216.72 &    1.81 &     0.07 \\
Aged 65+                                   &   1521.03 &   1517.41 &    0.09 &     0.93 \\
Average GCSE scores                        &    350.91 &    349.30 &    1.09 &     0.28 \\
Average PTAL score                         &      3.93 &      3.78 &    1.33 &     0.18 \\
Mean Income                                &  53676.62 &  52758.07 &    0.65 &     0.52 \\
Median House Price                         & 391187.98 & 375510.38 &    0.78 &     0.43 \\
No. of Full-time employees                 &   6179.66 &   5147.93 &    0.68 &     0.50 \\
No. of Part-time employees                 &   2206.21 &   1871.72 &    1.06 &     0.29 \\
No. of dwellings                             &   5639.12 &   5445.98 &    1.99 &     0.05 \\
No. of lakes \& rivers                      &      0.14 &      0.13 &    0.19 &     0.85 \\
No. of parks                               &      1.19 &      1.16 &    0.23 &     0.82 \\
No. of tourist locations                   &      1.31 &      1.67 &   -0.69 &     0.49 \\
Total crime rate                           &     98.77 &     90.65 &    0.94 &     0.35 \\
\hline
\end{tabular}
\label{table:confounder_balance}
\end{table}

\subsection{Confounder Balance}\label{subsec:Confounder Balance}
The matching algorithm is also evaluated for its effectiveness to pair units with high confounder similarity. Table \ref{table:confounder_balance} shows the average confounder values for units with higher and lower treatment levels in each pair for the year 2013 (other years also hold similar results). The independence t-test was performed to realise the significance of the confounder similarity. For every confounder variable, the null hypothesis that its average value in high treatment units of each pair is identical to the average value in low treatment units of each pair cannot be rejected (i.e., the confounder values are highly similar between paired units). Furthermore, the absolute value of the test statistic is less than 1 for most confounders suggesting a significant confounder similarity between the matched pairs of high and low treatment levels. 

\begin{figure*}[t!]
    \centering
    \includegraphics[scale=0.4]{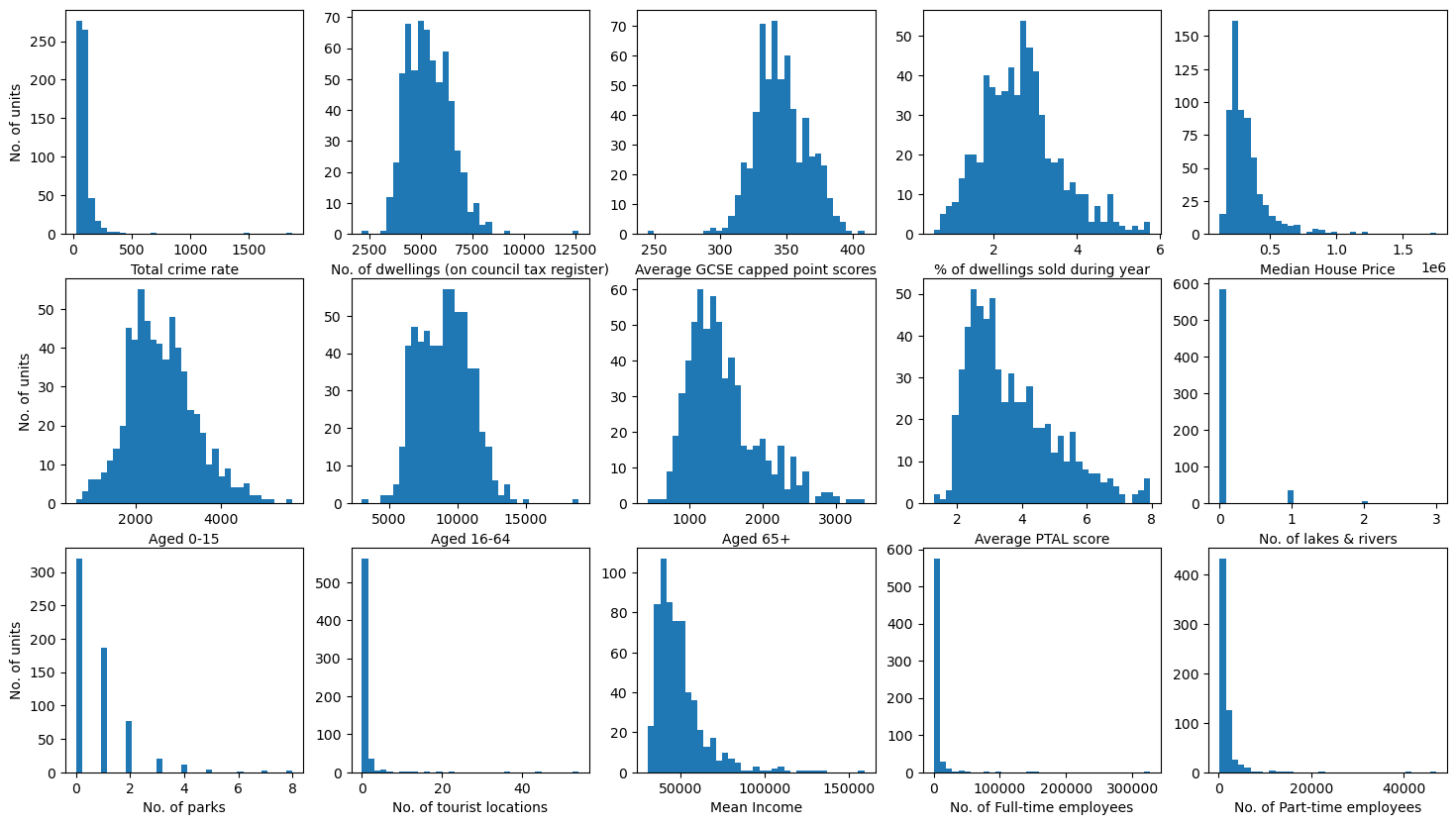}
    \caption{Confounder distributions from observational data}
    \label{fig:conf_dist}
\end{figure*}

\subsection{Treatment Effect}\label{subsec:Treatment Effect}
Now that it has been established that high treatment units and low treatment units have been sampled from almost identical populations, the average treatment effect is now computed. Table \ref{table:real_ATE} 
shows the average treatment effects calculated across all three years. The ATE, in this case, indicates the change in the number of tips per anchor store across all restaurants in the store's neighbourhood. An ATE value of zero implies no effect. Each anchor store in the neighbourhood caused an average increase between 0.05 and 0.15 tips to all restaurants in the neighbourhood. On average, this amounts to a 14.2-26.5\% increase in tips per restaurant between the matched neighbourhoods per year. Moreover, for all three years, ATE is positively skewed suggesting that anchor stores in a neighbourhood cause customer traffic to non-anchor restaurants. 

\begin{table}
\caption{Distribution of treatment effects across three years.}
\centering
\begin{tabular}{|l|r|r|r|r|r|}
\hline
year &  ATE &   min &   25\% &  75\% &   max \\
\hline
2011 & 0.13 & -5.10 & -0.50 & 0.75 &  9.53 \\
2012 & 0.19 & -3.33 & -0.33 & 0.56 &  4.95 \\
2013 & 0.16 & -5.60 & -0.38 & 0.58 & 35.48 \\
\hline
\end{tabular}

\label{table:real_ATE}
\vspace{-0.5cm}
\end{table}

\subsection{Synthetic Dataset}\label{subsec:Synthetic Dataset}
An effective way to evaluate the propensity score matching with known outcomes is to develop a synthetic dataset, following several works in this area
\cite{syn_data_ex, distance_metrics, Tsapeli_2017}. However, rather than developing an entirely random dataset, we generate the synthetic data with supervision from observed data. In essence, the influence of confounders on the treatment and the outcome as well as the treatment's influence on the outcome is modelled. Hence, the true causal effect of the treatment is known beforehand. Then, the matching method's efficacy to recover the true effect can be evaluated. 

To achieve this, we first generate $N$ units by generating confounders for each unit. The confounders are generated by modelling each confounder's distribution and sampling from it $N$ times.  Based on the distributions of the confounders from the observed data as shown in Figure \ref{fig:conf_dist}, it can be discerned that truncated normal distribution is used to model the confounders except for the number of full-time and part-time employees and the three tourism variables which follows the exponential distribution. Maximum likelihood estimation was utilised to determine the parameters of the distributions.

\begin{figure}[t!]
    \centering
    \includegraphics[scale=0.35]{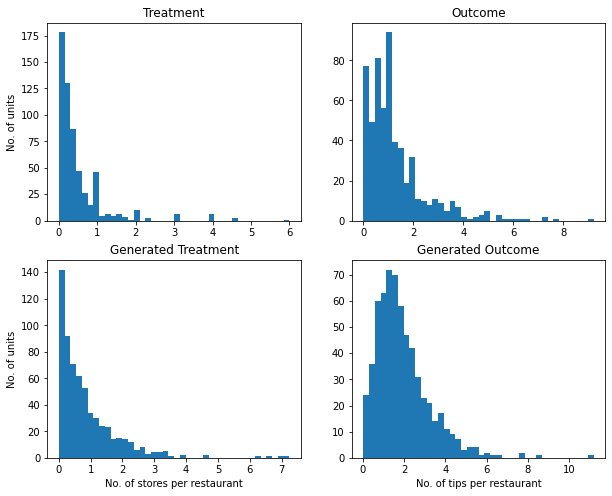}
    \caption{Top: true treatment and outcome distributions, Bottom: generated treatment and outcome distributions}
    \label{fig:treat_out}
    \vspace{-0.4cm}
\end{figure}

Then, two functions are modelled: $f_Z(x_{i})$ generates the treatment given the confounders, while $f_Y(z_{i}, x_{i})$ generates the outcome given the confounders and treatment. Based on the distributions of treatment and outcomes as depicted in figure~\ref{fig:treat_out}, exponential distribution can model the treatment as:

\begin{equation}
    f_{Z}(x) = P(Z = z|x_{i}) = 
    \begin{cases}
	    \frac{1}{f(x_{i})}e^{\frac{z}{f(x_{i})}}, & z \geq 0 \\
        0, &z < 0 \\
    \end{cases}
\end{equation}

\begin{equation}
    f(x_{i}) = \frac{\sum_{p}x_{ip}}{\eta}
\end{equation}

The truncated normal distribution is used to determine the outcome for each unit:

\begin{equation}
    f_{Y}(z_{i}, x_{i}) = Y_{i} \sim Norm\left( \alpha\sum_{p}x_{ip} + \beta \sum_{p}x_{ip} \right) + \gamma z_{i}
\end{equation}

$\alpha$, $\beta$ and $\eta$ are tuned until the range of values matches those of the observed data. $\gamma$ represents the value of the ground truth average treatment effect. Figure \ref{fig:treat_out} shows the treatment and outcome distributions for the observational and synthetic datasets are highly similar when these parameters are tuned. 

The synthetic data was generated for four values of average treatment effect. A total of 6250 units were generated for each treatment effect, while the matching was run on 10 batches of 625 units. The treatment effects values include large and small positive and negative values. Table \ref{table:synthetic_ATE} summarises the results.
The estimated ATEs from synthetic data are highly similar to the ground truth ATEs. While the treatment effects take a range of values across all cases, at least 75\% of effects are positive effects in most cases. This is consistent with the results obtained from the real data. We also observe that the matching performs well on larger ground truth ATE compared to the lower values as the difference between true and estimated ATE is large for smaller ATE values. This suggests that one must take a cautious stance when interpreting the results with real data in Table 6.

\begin{table}[t!]
\caption{Treatment effects for synthetic data generated using various ground truth average treatment effects. }
\centering
\begin{tabular}{|r|r|r|r|r|r|}
\hline
   ATE &  est. ATE &     min &    25\% &    75\% &      max \\
\hline
 0.25 &     0.22 & -57.11 & -0.84 &  1.22 &   48.52 \\
 1.50 &     2.06 & -74.76 &  0.52 &  2.57 & 1430.55 \\
 4.00 &     4.14 & -41.69 &  3.03 &  5.12 &  170.27 \\
10.00 &     9.80 & -71.70 &  8.92 & 11.02 &   60.46 \\
\hline
\end{tabular}
\label{table:synthetic_ATE}
\vspace{-0.5cm}
\end{table}

Based on the results from correlation analysis and matching, we observe that neighbourhood characteristics have a significant impact on the presence of anchor stores and their subsequent causation of increased customer visitation to non-anchor stores. The positive treatment effects observed across all three years reinforce previous work in \textit{Economics} on the anchor store effect. This consistency in results with the literature and the performance across the presented evaluation metrics reveal the viability of applying causal inference to investigations delving into the retail industry.


\section{Discussion and Future Work}\label{sec:Discussion}

The above findings lead us to discuss the implications and provide practical application scenarios for researchers, small business owners, mall developers, and mall managers. A few examples are as follows.
\begin{enumerate}

    \item[($\mathcal{I}_1$)] Non-anchor stores can utilise the causal analysis for site selection decisions by targeting neighbourhoods with significant anchor store effects. For instance, they can compare the customer traffic between current and new locations based on the anchor store presence. 
    
    \item[($\mathcal{I}_2$)] Similarly, The low dose units in matched pairs with a dose difference serve as future location recommendations to anchor stores and mall developers as they demonstrate similar neighbourhood conditions to the high dose counterparts with lower competition.
    
    \item[($\mathcal{I}_3$)] Mall managers can derive the value of their malls from the anchor store effects in their neighbourhoods and develop lucrative rental contracts for their tenants.

\end{enumerate}

Nonetheless, there are limitations in our work which must be mentioned along with potential future works.

\textbf{Unknown Confounding Variables.} 
In our analysis, we took into account 15 confounding variables of 6 different aspects. Nevertheless, there may still exist several variables with confounding effects for example various social events that impact customer traffic to anchor and non-anchor stores alike. As pointed out in~\cite{hasthanasombat_understanding_2019}, uncertainty pertains to whether all confounding variables have been explored. A systematic approach to solving this challenge remains an open problem.

\textbf{Data Availability.}
The dataset studied is bottlenecked by time and space due to the limited data availability. The census data below annual granularity and for other cities has been challenging to find. Also, the dataset only covers three years, limiting our analysis to the period from 2011 to 2013. The geography and time in our study are discretised by neighbourhood and year respectively, which might be too broad. 
It is worthwhile to employ causal modelling like Granger Causality~\cite{granger_causality} to leverage spatiotemporal properties of LBSNs and study this effect at a more fine-grained level~\cite{vallapuram_interpretable_asonam21,chen_ian_asonam21}.

\section{Conclusion}
\label{sec:Conclusions}

In this paper, we adopted a causal approach and quantitatively demonstrated the anchor store effect. Contrary to previous work in this area that perform studies on confidential datasets, we employed publicly available data from an LBSN and a government’s data-sharing portal making our study more cost-effective. Additionally, we employed a causal analysis framework based on propensity score matching to mitigate the bias induced by confounding effects between variables that impact customer traffic to anchor stores and non-anchor stores. As a case study, we applied this framework to over 600 neighbourhoods in the Greater London area and concluded that the anchor stores cause an increase of 14.2-26.5\% tips per year to non-anchor stores within their neighbourhoods. Furthermore, we performed extensive evaluations that demonstrated the framework's high matching quality and resilience to noise through experiments on synthetic datasets with known treatment effects.  

Though we have realised a causal relation in its simplest form between two variables (i.e. treatment and outcome), we hope that further investigations will be undertaken to discern more complex causal mechanisms in Retail Economics. 

\section*{Acknowledgments}\label{sec:Acknowledgement}
This research has been supported in part by the 5GEAR project (Decision No. 318927) and the FIT project (Decision No. 325570) from the Academy of Finland. 

\bibliographystyle{IEEEtran}
\bibliography{ref}

\end{document}